\documentclass[prb,aps,twocolumn,showpacs,superscriptaddress]{revtex4}

\usepackage{amsmath,amsfonts}
\usepackage{graphicx}

\usepackage{dcolumn}
\usepackage{longtable}
\usepackage{color}

\begin{document}

\title{ Effect of charge doping on the electronic structure, orbital
  polarization, and structural distortion in nickelate superlattice}

\author{Heung-Sik Kim}
\affiliation{Department of Physics, Korea Advanced Institute of Science and Technology, 
Daejeon 305-701, Korea}

\author{Myung Joon Han}
\affiliation{Department of Physics, Korea Advanced Institute of Science and Technology, 
Daejeon 305-701, Korea}
\affiliation{KAIST Institute for the NanoCentury, KAIST, Daejeon 305-701, Korea}

\begin{abstract}
Using first-principles density functional theory calculations, we
investigated the effect of charge doping in a LaNiO$_3$/SrTiO$_3$
superlattice. The detailed analysis based on two different doping simulation
methods clearly shows that the electronic and structural
properties change in a systematic way, and that the orbital polarization
({\it i.e.} relative occupation of two Ni-$e_g$ orbitals) is reduced
and the Ni to apical oxygen distance is enlarged as the number of doped
electrons increases.  Also, the rotation angles of the NiO$_6$/TiO$_6$
octahedra strongly and systematically depend on the doping.  Our
results not only suggest a possible way to control the orbital and
structural properties by means of charge doping, but also provide useful
information for understanding experiments under various doping
situations such as oxygen vacancy.
\end{abstract}

\pacs{ 75.25.Dk, 75.70.Cn, 73.21.Cd}

\maketitle

\section{Introduction}

Recent technical advances in the atomic-scale growth of transition
metal oxide (TMO) heterostructures have created considerable research
interest \cite{MRS, Hwang_review}. In TMO heterostructures, multiple degrees of
freedom ({\it i.e.}, charge, spin, orbital, lattice) are coupled to
each other, often creating novel material characteristics such as
high-temperature superconductivity and colossal magneto resistence
\cite{MIT-RMP}. 
Using atomic-scale techniques to fabricate artificial TMO heterostructures
makes it possible to control those degrees of freedom and band structures, and
therefore to create or design new `correlated electron' properties.
Previous TMO superlattice studies
\cite{Ohtomo-1,Okamoto,Ohtomo-2,Nakagawa,OrbReconst} have shown that
many unexpected material phenomena can be realized at the TMO
heterointerface, such as magnetism and superconductivity
\cite{mag-2007,mag-LAOSTO-1,mag-LAOSTO-2,Reyren,Gibert}.

When combined with other degrees of freedom, charge doping can play
a significant role in determining material properties of TMO heterostructures.
Sometimes extra charges are introduced in an unexpected and uncontrolled way.
For example, oxygen vacancies often drives a TMO system to exhibit fairly
different material characteristics ({\it e.g.,} driving an insulating material
to be metallic) \cite{O-vac1,O-vac2,O-vac3}.  Cation inter-mixing can
also be important, as it introduces a different local ionic potential
to the nearby atoms ({\it e.g.,} the inter-mixing of Sr$^{2+}$ and
La$^{3+}$). It is also possible to control the
amount of extra charges by chemical doping or electric field, for
example.  Considering all these possibilities, it is important to
understand the effect of charge doping in TMO heterostructures.
In particular, the relation between the rotation of metal-oxygen
octahedra and charge doping has never been investigated in a
systematic way.

Among the various TMOs, rare-earth nickelate compounds $R$NiO$_3$,
where $R$ denotes rare-earth elements, show interesting properties.
The $R$NiO$_3$ series exhibit a range of transitions,
from the ``correlated metal'' phase of LaNiO$_3$ 
to charge-ordered insulating states as the ionic $R$ becomes smaller
and more octahedral distortion is introduced\cite{RNO_review}. 
In such systems, the systematic dependence of the transition 
temperature $T_{\rm MI}$
with respect to the ionic size of $R$ implies strong coupling
between the structural and electronic degrees of freedom. 
In this regard, introducing reduced dimensionality through the synthesis of
nickelate compound heterostructures can be another promising 
way to control the material properties and realize emergent physics 
in this system\cite{Chaloupka,Hansmann,LNOLAO-NatMatt}.
Indeed, it has recently been reported that LaNiO$_3$,  
which is a paramagnetic metal in the bulk phase, 
shows insulating behavior with magnetic order in a 
superlattice\cite{Boris-Science,Frano}.
There have been a number of subsequent studies about nickelate 
heterostructures\cite{Kaiser-PRL,Chakhalian,May,NNO-PRL,Stemmer-new,
Frano,Pentcheva,SIC,MJHan-DMFT,MJHan-opol,MJHan-LDAU,MJHan-LNOSTO}, 
which show a potential for rich physics, but there has not been 
systematic studies about the effect of doping in these systems yet.

In this paper, we examine the effect of doping on the nickelate
superlattice by taking (LaNiO$_3$)$_1$/(SrTiO$_3$)$_1$ as a prototype example
\cite{Kaiser-PRL,Son,Stemmer-new,MJHan-LNOSTO}. 
We employed first-principles density functional theory (DFT) 
calculations augmented with mean-field treatment of on-site Coulomb
interaction, the so-called DFT+$U$ methodology, to consider the role 
of electron correlations inherent in the Ni $e_{\rm g}$ orbitals.
The behavior of the electronic structure, orbital polarization, magnetism,
and structural distortion with respect to the amount of charge doping is
examined, and the effect of Coulomb interaction is also discussed. 
We pay our special
attention to the orbital order within the Ni $e_{\rm g}$ states and
the structural responses to it. Two different structural settings have
been considered: (i) the calculations in which only tetragonal
distortions of NiO$_6$ octahedra (elongation or contration of Ni--O
bond length along the normal direction to the superlattice plane) are
allowed and (ii) the full NiO$_6$ octahedral distorion including
rotation and Jahn-Teller distortion.  In the first set, we investigate
how the doping-induced change of orbital polarization affects the
NiO$_6$ octahedra. The relation between the octahedral rotation
pattern and the local distortion is examined with the second sets.
Hereafter we call the first and the second sets as the
`tetragonally-distorted' and `fully-distorted' cells, respectively.
Systematic changes are found: As electrons are introduced, the orbital
polarization ({\it i.e.,} the relative occupation of $d_{x^2-y^2}$ to
$d_{3z^2-r^2}$) is reduced, and the Ni to apical oxygen distance is
enhanced.  The evolution of the octahedral rotation angles also
exhibits a systematic tendency on charge doping. Our results suggest a
possibility to control the structural property as well as the
electronic structure by doping, and provides useful information to
understand the experiments under the various types of doping
situations.

After presenting the computational details in Sec.~\ref{sec2}, we present and
discuss the calculated phase diagrams in
Sec.~\ref{sec-PD}.  We further elaborate our results of the
tetragonally (Sec.~\ref{sec3}) and fully distorted (Sec.\ref{sec4})
calculations in the following sections.
In consideration of recent theoretical and experimental results on
the relevant systems, we discuss the implications and limitations of our
results.  Further issues are discussed in Sec.~\ref{sec5}, and we
summarize our results in Sec.~\ref{sec6}.

\section{Computation Details}
\label{sec2}

For the electronic structure calculations, we used DFT within local
spin density approximation (LSDA)\cite{CA} and the projector-augmented
wave (PAW) method \cite{PAW} as implemented in the Vienna {\it ab
  initio} simulation package \cite{VASP}. We adopted a plane-wave
energy cutoff of 400 eV with a $6\times 6\times 4$ {\it k}-point
sampling on the Monkhorst-Pack grid. To incorporate the effect of
electron correlations within the Ni $d$ orbital, the so-called
simplified version of rotationally invariant DFT+$U$ as suggested by
Dudarev {\it et al.} \cite{Dudarev} was used with the effective
$U_{\rm eff}\equiv U-J$ varying from 0 to 5 eV.  In the previous
literature, the relatively small value of $U\sim$3 eV has been taken as a
reasonable value \cite{Chakhalian,Gibert,ATLee} for a nickelate
superlattice although it can be regarded as too small. Since the value
of $U_{\rm eff}$ can be an important issue, we scanned the range of
$U_{\rm eff}$ and discuss
its effect on the electronic and magnetic structure. For each doping
and $U_{\rm eff}$ value, the structural optimization was performed
with a force criterion of 1 meV/\AA~.  During this process, a
ferromagnetic (FM) and a checkerboard-type antiferromagnetic (AF)
order were set as the initial magnetic configurations.

\begin{figure}
  \centering
  \includegraphics[width=0.45\textwidth]{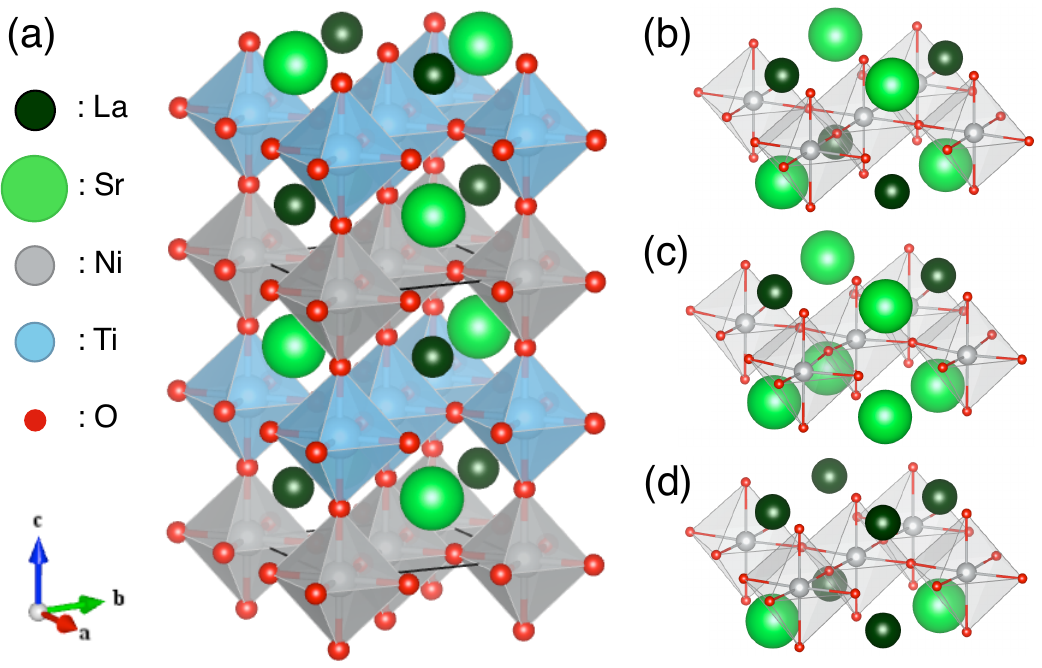}
  \caption{(Color online) (a) The atomic structures used in the
    calculations. The black thin lines show the unit cell.  (b)-(d)
    The arrangement of La and Sr ions are shown in (b)
    (LaNiO$_3$)$_1$/(SrTiO$_3$)$_1$, (c)
    (La$_{0.5}$NiO$_3$)$_1$/(Sr$_{1.5}$TiO$_3$)$_1$, (LaSr$_3$-cell)
    and (d) (La$_{1.5}$NiO$_3$)$_1$/(Sr$_{0.5}$TiO$_3$)$_1$
    (La$_3$Sr-cell), respectively.}
  \label{fig:struct}
\end{figure}

Fig.~\ref{fig:struct} shows the unit cell structure of the
(LaNiO$_3$)$_1$/(SrTiO$_3$)$_1$ superlattice used in this study. As
shown in Fig.~\ref{fig:struct}(a) and (b), we chose
a $\sqrt{2}\times\sqrt{2}$ supercell with La-Sr intersite mixing to
avoid the internal electric fields.  For the in-plane lattice constant
we used 3.905$\times\sqrt{2}$~\AA, corresponding to the SrTiO$_3$
value. The out-of-plane lattice constant is optimized for each doping
level and $U_{\rm eff}$ value. For a subset of doping and $U_{\rm
  eff}$ values ($U_{\rm eff}$ = 0, 3, and 5 eV; $\pm$0.5 and 0
electrons per Ni) we performed the $2 \times 2$ supercell calculations
(four Ni atoms per unit cell) and obtained consistent results with
the $\sqrt{2}\times\sqrt{2}$ unit cell.

To investigate the influence of charge doping we mainly
used the rigid band shift method, in which the doped charges are
compensated by the uniform background of the opposite sign. Seven
different charge dopings have been considered: $\pm$0.50, $\pm$0.25,
$\pm$0.125, and 0.00 electrons per Ni atom (hereafter denoted as $\pm$0.50$e$,
$\pm$0.25$e$, $\pm$0.125$e$, and the undoped cell, respectively).  
A different type of doping was also considered by changing the Sr/La ratio. Since our
unit cells contain four $A$-site cations, $\pm$0.50$e$ of
doping can be simulated with the case of
(La$_{0.5}$NiO$_3$)$_1$/(Sr$_{1.5}$TiO$_3$)$_1$, and
(La$_{1.5}$NiO$_3$)$_1$/(Sr$_{0.5}$TiO$_3$)$_1$ as shown in
Fig.~\ref{fig:struct}(c) and (d) respectively. These two will be
hearafter denoted as LaSr$_3$- and La$_3$Sr-cell, respectively. 

The orbital polarization, representing the relative occupations in the
two Ni-$e_g$ orbitals, can be defined as\cite{MJHan-opol}:
\begin{align}
P_{e_g} &= \frac{n_{d_{x^2-y^2}} - n_{d_{3z^2-r^2}}}{n_{d_{x^2-y^2}} + n_{d_{3z^2-r^2}}}, \\
n_i    &= \int^{\epsilon_{\rm F}}_{\epsilon_{\rm b}} f_i(\epsilon) d\epsilon,
\label{eq:orbpol}
\end{align}
where we choose $\epsilon_{\rm b} = -3.5$ eV (with the Fermi energy
$\epsilon_{\rm F} = 0$) to capture the occupations on 
the valence $d$-orbital-like states. We found that our conclusions
were unchanged even when we used the values of $\epsilon_{\rm b}$ down
to $-10$ eV.

\section{Overall phase diagram}
\label{sec-PD}

\begin{figure}
  \centering
  \includegraphics[width=0.45\textwidth]{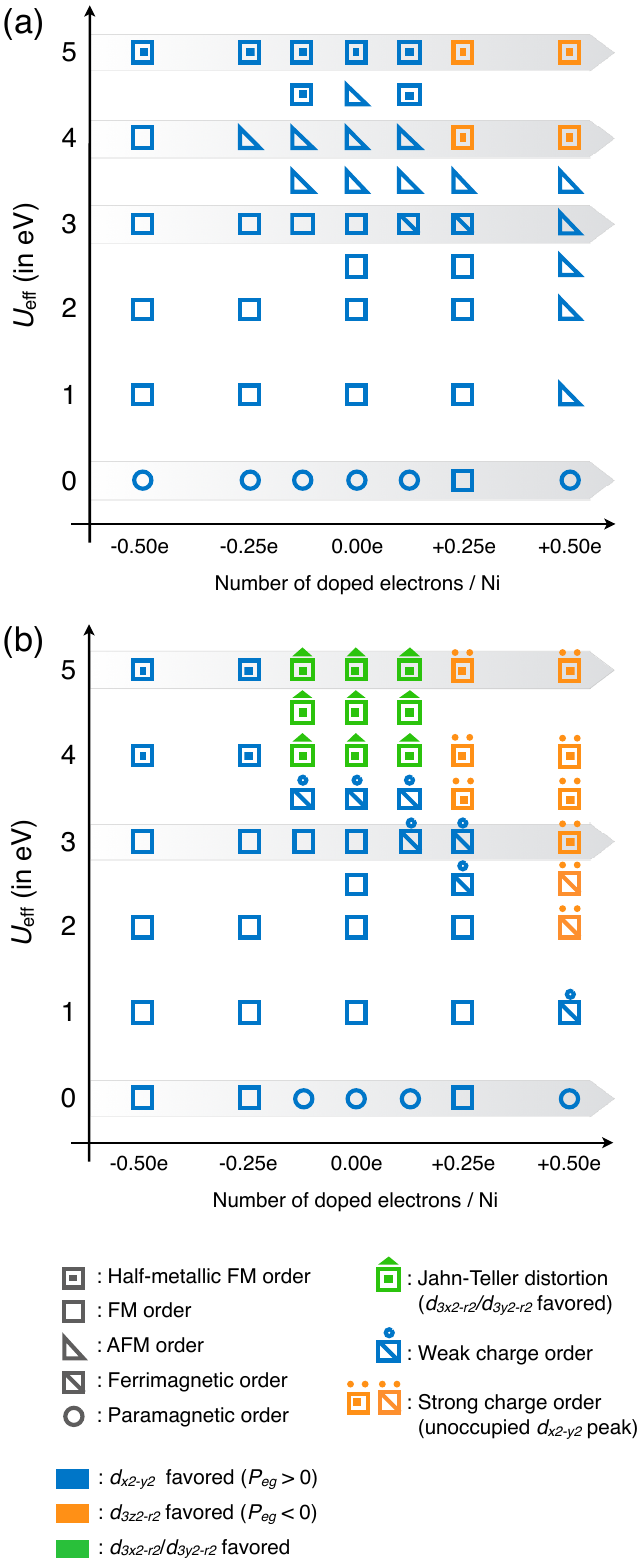}
  \caption{(Color online) Phase diagrams of (a) the tetragonally and
    (b) the fully distorted calculations in the doping -- $U_{\rm
      eff}$ space. The shape and color of the symbols denote the types
    of spin and orbital orders, respectively. In (b), the type of
    NiO$_6$ octahedral distortions are denote d by the additional
    symbols of small triangle or circle. The shaded area represents
    the parameter regions we mainly focus on in this study.  }
  \label{fig:PD}
\end{figure}

Let us first take an overall view of our results in the parameter
space spanned by charge doping and $U_{\rm eff}$.
Fig.~\ref{fig:PD}(a) and (b) show the results from the tetragonally
and fully distorted calculations, respectively.  The shape and the
color of the symbols denote the type of the magnetic order and the
orbital polarization, respectively.  Comparing Fig.~\ref{fig:PD}(a) to
(b), one can notice the following points. i) Due to the confinement
effect, $d_{x^2-y^2}$ is preferred over the $d_{3z^2-r^2}$ orbital in the
majority region of the parameter space, while in the electron-doped
and the large $U_{\rm eff}$ regions, the $d_{3z^2-r^2}$ orbital is
favored and $P_{e_{\rm g}}$ becomes negative.  The area of the
negative $P_{e_{\rm g}}$ region is significantly enlarged as the
NiO$_6$ octahedra distortion is fully allowed (Fig.~\ref{fig:PD}(b)),
which is indicative of the cooperation between the orbital and the structural
degrees of freedom in these compounds.  ii) In the small-$U_{\rm eff}$ ($\lesssim$ 2 eV)
regime the paramagnetic or weak ferromagnetic metal phase is stable,
while the `half-metallic' ferromagnetic phase with gapless majority and
gapped minority spin channel is stabilized when $U_{\rm eff}$ is large 
($\gtrsim$ 4 eV).  
The preference for the
ferromagnetic phase is consistent with a previous DFT+$U$ study in
these regimes \cite{Pentcheva,May,MJHan-LDAU}, although these results
are not well compared with the recent experimental observations that
report the insulating phase with no magnetization in ultrathin
LaNiO$_3$ heterostructures\cite{Boris-Science,Frano}.
iii) In
between the weak and strong $U_{\rm eff}$ regime ($U_{\rm eff}\simeq$
3 eV), the metallic phase with antiferromagnetic order (ferrimagnetic
order in the fully distorted calculations) is preferred over the
ferromagnetic states.  The antiferromagnetic phases in the
tetrogonally distorted cells were not captured by the previous
DFT+$U$ study, in which sparse sets of $U_{\rm eff}$ values 
were used\cite{Pentcheva,MJHan-LDAU}.
The antiferromagnetic phase turns into ferrimagnetic phases in the
fully distorted calculations. This type of spin order was also not
observed in previous studies since these phases become favorable only
in the narrow area in the parameter space as shown in Fig.~\ref{fig:PD}(b)
\cite{Pentcheva,May}.  Since it is argued that DFT+$U$ is inconsistent
with experiments in terms of FM spin ground state\cite{SIC}
it is important to note this point.  iv) In the fully distorted cells,
a strong Jahn-Teller distortion exists near the undoped regime
with $U_{\rm eff}$ larger than 4 eV.  On the other hand, the NiO$_6$
breathing mode having two inequivalent Ni sites, reported in other
DFT+$U$ studies \cite{Pentcheva,Chakhalian}, was not observed in our
undoped cells
partly due to the difference in the cation ordering.
Instead, we found the breathing mode in the electron-doped regime with
$U_{\rm eff}$ larger than 2 eV.

\section{Results with tetragonal distortion}
\label{sec3}

In this section we present the calculation results of the tetragonally
distorted cells.  The results are meaningful for understanding the system
in further details and may be relevant to experimental situations
in which the rotational distortion modes are suppressed for some
reason, although a recent study of LNO/STO indicates the possible
rotations \cite{Stemmer-new}. For simplicity, we focus on the shaded
regions of the phase diagram in Fig.~\ref{fig:PD}(a).

\subsection{Structural changes}

\begin{figure}
  \centering
  \includegraphics[width=0.47\textwidth]{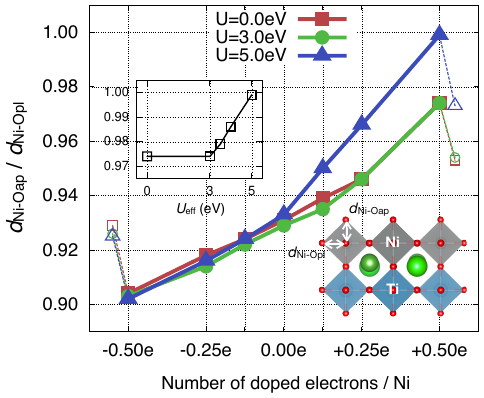}
  \caption{(Color online) The calculated ratio of $d_{\rm Ni-O_{ap}} /
    d_{\rm Ni-O_{pl}}$ as a function of doping. The $d_{\rm
      Ni-O_{ap}}$ and $d_{\rm Ni-O_{pl}}$ are illustrated in the lower
    right inset. The filled and empty symbols indicate the results
    from the rigid band shift and the cation substitution calculations,
    respectively. The upper left inset shows the $d_{\rm Ni-O_{ap}} /
    d_{\rm Ni-O_{pl}}$ ratio as a function of $U_{\rm eff}$ in the
    case of +0.50$e$ cell.}
  \label{fig:atomdisp}
\end{figure}

The most significant structural parameter changed by doping is the
Ni to apical oxygen distance ($d_{\rm Ni-O_{ap}}$).
Figure \ref{fig:atomdisp} shows the ratio between $d_{\rm Ni-O_{ap}}$
and the Ni to planar oxygen distance ($d_{\rm
  Ni-O_{pl}}$), which is fixed in the tetragonally
distorted cells.  Increasing trends with respect to electron doping is
clearly seen. The result of 
the LaSr$_3$/La$_3$Sr-cell calculations (empty symbols) is consistent with
the rigid band shift calculations (filled symbols).

In most cases, the $d_{\rm Ni-O_{ap}} / d_{\rm Ni-O_{pl}}$ is below the unity
due to the tensile strain.  Increasing $U_{\rm eff}$, however,
enhances $d_{\rm Ni-O_{ap}} / d_{\rm Ni-O_{pl}}$ in the electron-doped
regime, as can be seen in the inset of Fig.~\ref{fig:atomdisp}.  The
abrupt change of $d_{\rm Ni-O_{ap}} / d_{\rm Ni-O_{pl}}$ can likely be
attributted to the $U_{\rm eff}$-dependence of
$d_{x^2-y^2}$/$d_{3z^2-r^2}$-orbital occupations which will be
discussed further in the following subsection.

\subsection{Electronic structure, orbital polarization, and magnetism}

\begin{figure}
  \centering
  \includegraphics[width=0.47\textwidth]{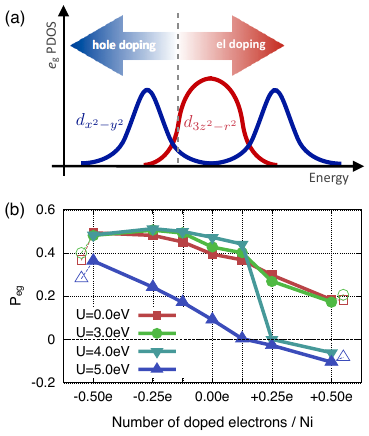}
  \caption{(Color online) (a) Schematic PDOS of Ni $d_{x^2-y^2}$- and
    $d_{3z^2-r^2}$-orbital. (b) The calculated $e_{\rm g}$-orbital
    polarization as a function of doping where the filled and empty
    symbols correspond to the results from rigid band shift and
    LaSr$_3$/La$_3$Sr-cells calculations, respectively.}
  \label{fig:orbitpol}
\end{figure}

The doped charges reside mostly in the Ni $e_{\rm g}$ bands, since the
Ti-$t_{2g}$ states are well separated from the Fermi level and the
Ni-$t_{\rm 2g}$ are fully filled, in all configurations except the
tetragonally distorted cells with $U_{\rm eff}=$ 0 eV.  In the case of
La$_3$Sr- or the LaSr$_3$-cells, a small fraction of Ti-$t_{\rm 2g}$
bands touches the Fermi level at $U_{\rm eff}$=0, and they are pushed away from
the Fermi level as $U_{\rm eff}$ is included. We also note that all
of the results remain metallic regardless of the value of $U_{\rm eff}$, 
doping, and the presence of octahedral rotation or distortion.

From the shape of Ni $e_{\rm g}$-projected density of states (PDOS),
schematically shown in Fig.~\ref{fig:orbitpol}(a), it is expected that
the major portion of the doped charges fill up the $d_{3z^2-r^2}$
orbitals and $P_{\rm e_g}$ changes accordingly (See
Fig.~\ref{figA:pdos_d} and \ref{figA:pdos_eg} in the Appendix for the full
details of PDOS).  Fig.~\ref{fig:orbitpol}(b) shows that while the
orbital polarization decreases as the more electrons are doped over
the entire range of $U_{\rm eff}$, the detailed behavior is different,
and depends on the $U_{\rm eff}$ value. In the region of $U_{\rm
  eff}\leq$ 3 eV, the $P_{\rm e_g}$ decreases gradually as electrons
are doped.  For $U_{\rm eff}\geq$ 5 eV, $P_{\rm e_g}$ is reduced
significantly and changes sign when more than +0.25$e$ of electons
per Ni are doped.  The results of $U_{\rm eff}=$ 4 eV is in between
the two, and exhibits the behaviors of both regimes. These features
are consistent with the behavior of $d_{\rm Ni-O_{ap}} / d_{\rm
  Ni-O_{pl}}$ under electron doping; more occupations in
$d_{3z^2-r^2}$ ({\it i.e.,} the smaller $P_{e_g}$) should favor longer
$d_{\rm Ni-O_{ap}}$ owing to the Coulomb repulsion between the
electrons occupying the Ni $d_{3z^2-r^2}$ orbital and the $p_z$
orbital in the apical oxygen.  Accordingly, NiO$_6$ octahedra are
elongated along the interface-normal direction as electrons are
introduced. Note that the La$_3$Sr- or the LaSr$_3$-cell calculations
(open symbols in Fig.~\ref{fig:orbitpol}(b)) shows results
consistent with the rigid band shift calculations.

\begin{figure}
  \centering
  \includegraphics[width=0.48\textwidth]{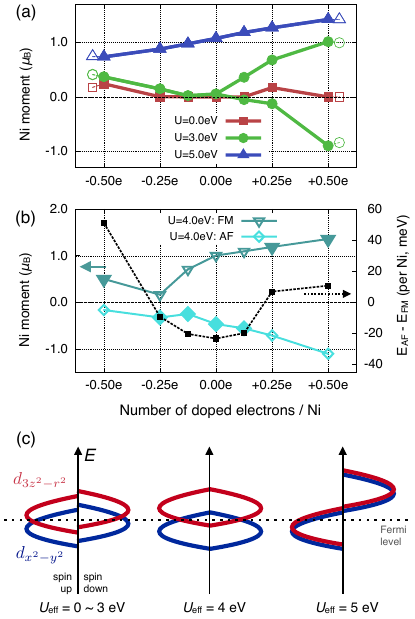}
  \caption{(Color online) (a) The calculated Ni moments as a function
    of doping with $U_{\rm eff}=$ 0, 3, and 5 eV. The branching of the
    $U_{\rm eff}=$ 3 eV curve (green color) reflects the two
    inequivalent Ni sites in the unit cell.  (b) Size of Ni moments in
    the ferromagnetically and antiferromagnetically (depicted by
    negative sign) ordered states, and the energy difference between
    the two phases.  The moment size of the energetically favorable
    state is depicted by the filled symbols and the other by the
    open symbols. (c) Schematic shape of undoped Ni $e_{\rm g}$ PDOS
    at $U_{\rm eff}=$ 0--3, 4, and 5 eV.  }
  \label{fig:mag}
\end{figure}

Figure \ref{fig:mag}(a) shows the size of the Ni magnetic moment as a
function of doping at $U_{\rm eff}=$ 0, 3, and 5 eV. When $U_{\rm
  eff}$ is less than 3 eV, the hole doping induces ferromagnetic
moments. The schematic shape of PDOS suggests that, in this range of
$U_{\rm eff}$, the Stoner mechanism may be responsible for it in
hole-doped systems.  On the other hands, the antiferromagnetic phase
is stabilized in $3 \leq U_{\rm eff} \leq 4.5$ eV (see
Fig.~\ref{fig:PD}(a)).  In between the ferro- and the
antiferromagnetic phases, a ferrimagnetic order is induced in the
electron doped regime at $U_{\rm eff}=$ 3 eV. 
The relative energy between the
converged ferromagnetic and antiferromagnetic phases at $U_{\rm eff}=$
4 eV, as well as the size of the moments, are shown in
Fig.~\ref{fig:mag}(b).  One can notice two crossovers as the system
evolves from the hole-doped to the electron-doped regime; the
crossover from a ferromagnetic to an antiferromagnetic phase around
$-$0.25$e$ doping, and from the antiferromagnetic to a half-metallic
phase after +0.25$e$ doping.  The half-metallic phase, which exibits a
small value of $P_{\rm e_g}$ as depicted in
Fig.~\ref{fig:orbitpol}(b), extends to the entire range of doping
above $U_{\rm eff}=$ 5 eV.

The type of magnetism and the size of orbital
polarization shows an interesting relation.  The weakly correlated
ferromangetic and the antiferromagnetic phase exibit a larger value of
$P_{\rm e_g}$, while the half-metallic phase favors a small and even
negative $P_{\rm e_g}$. The situation is schematically depicted in
Fig.~\ref{fig:mag}(c); In the weakly correlated ferromagnetic phase
the confinement effect favors the $d_{x^2-y^2}$ orbital rather than
$d_{3z^2-r^2}$, and at $U_{\rm eff}=$ 4 eV the orbital polarization
increases and antiferromangetic phase originating from the
$d_{x^2-y^2}$-like states occurs.  Such tendency can be seen in
Fig.~\ref{fig:orbitpol}(b), where $P_{\rm e_g}$ is enhanced as $U_{\rm
  eff}$ becomes larger up to 4 eV near the undoped regime. On the
contrary, as $U_{\rm eff}$ goes beyond that ($U_{\rm eff}$=5 eV),
driving the system half-metallic becomes more favorable owing to the large
Ni $e_{\rm g}$ bandwidth, and the $P_{\rm e_g}$ drops significantly.
One can speculate that, up to a moderate strength, $U_{\rm eff}$ tries
to lower the energy by enhancing the orbital polarization and driving
the system antiferromagnetic. Beyond that, a half-metallic system with
less orbital polarization is energetically favorable (see
Fig.~\ref{fig:mag}(c)).

It should be noted that the origin of the spurious half-metallic ferromagnetic
phase is the energy shift of the Ni $e_{\rm g}$ bands induced by electron
correlations, which was the only way that DFT+$U$ methodology could 
reduce the Coulomb energy in this system as depicted in 
Fig.~\ref{fig:mag}(c). 
More elaborate treatment of electron correlations, however, can take 
the bandwidth renormalization effect into account\cite{HPark,MJ-GW}, 
which can facilitate the orbital ordering scenario 
(compare the center and right figures of Fig.~\ref{fig:mag}(c)).
Also, although the presence of a half-metallic 
ferromagnetic phase seems robust in the strong $U_{\rm eff}$ regime, 
it suffers from a large amount of ferromagnetic moments 
which is inconsistent with experimental observations of no net
magnetic moments in nickelate superlattices\cite{Boris-Science,Frano}. 
In this regard, we speculate that the half-metallic 
ferromagnetic phase can be considered as an artifact of an inappropriately
large value of $U_{\rm eff}$ for Ni $e_{\rm g}$ orbital in the DFT+$U$ 
formalism.

\section{Results with octahedral rotation}
\label{sec4}

In this section, the full octahedral distortions are taken into
account in combination with the effect of doping. As a
starting configuration for the structural optimization, we considered
four different cases of $(a^0a^0c^-)$-, $(a^-a^-c^0)$-, and
$(a^-a^-c^\pm)$-type rotations (following the Glazer notation as defined
in Ref.~\onlinecite{Glazer,May,Rondinelli-Spaldin-2}, where the
positive and negative signs refer to the ferro- and
antiferro-distortive rotation, respectively) and performed the
structure relaxations with the symmetry enforced. The structures
obtained by this process were further optimized with the symmetry
constraint turned off.  Without the constraint, the oxygen cage shape
can deviate from the ideal one. Therefore, we present the averaged
values of the bond lengths and the rotation angles.  Also, with
the removal of the structural constraints, the Jahn-Teller- or
breathing type distortion of NiO$_6$ octahedra is obtained in some
cases.
For simplicity, we focus on the shaded
regions of the phase diagram in Fig.~\ref{fig:PD}(a).

\subsection{Structural changes}

\begin{figure}
  \centering
  \includegraphics[width=0.48\textwidth]{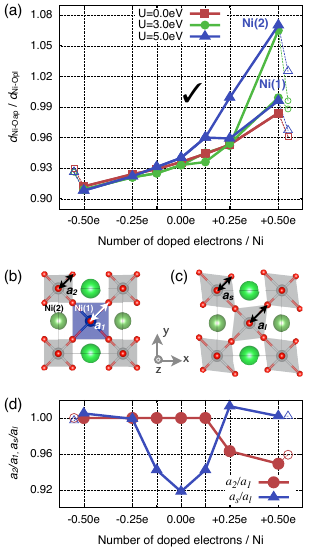}
  \caption{(Color online) (a) The calculated $d_{\rm Ni-O_{ap}} /
    d_{\rm Ni-O_{pl}}$ as a function of doping in the fully distorted
    calculations. The branching of the $U_{\rm eff}=$ 3 and 5 curves
    in the electron-doped regime reflects the two inequivalent sites,
    Ni(1) and Ni(2). (b) Schematic figures for the breathing type and
    (c) the Jahn-Teller type distortion.  In (b), the planar Ni-O bond
    length of Ni(1) and Ni(2) are denoted as $a_1$ and $a_2$,
    respectively. In (c), the longer and shorter Ni-O bond lengths in
    the Jahn-Teller-distorted octahedra are denoted by $a_l$ and
    $a_s$, respectively.  (d) The amount of distortions at $U_{\rm
      eff}=$ 5 eV are represented by the ratio of $a_2$/$a_1$ and
    $a_s$/$a_l$.  }
  \label{fig:rot_local}
\end{figure}

The evolution of the Ni-O bond length ratio $d_{\rm Ni-O_{ap}} /
d_{\rm Ni-O_{pl}}$ in the fully distorted calculations are shown in
Fig.~\ref{fig:rot_local}(a) as a function of doping.  An increasing
trend of $d_{\rm Ni-O_{ap}} / d_{\rm Ni-O_{pl}}$ is obtained, which is
mainly attributed to the enlarged $d_{\rm Ni-O_{ap}}$.  Even in the
presence of the octahedral rotation the $d_{\rm Ni-O_{pl}}$ is not
significantly changed unless any additional distortion is introduced
(as will be discussed further below).  Due to the breathing-type
distortion of the two inequivalent NiO$_6$ octahedra, the ratio of
$d_{\rm Ni-O_{ap}} / d_{\rm Ni-O_{pl}}$ is branched into two parts in
the electron-doped regime at $U_{\rm eff}=$3 and 5 eV.  
The $d_{\rm Ni-O_{ap}} / d_{\rm Ni-O_{pl}}$ is
significantly larger than the values for the tetragonally distorted
case in Fig.~\ref{fig:atomdisp}, because the octahedral rotation
creates more room for the bond length elongation.  This observation
indicates the stronger preference to $d_{3z^2-r^2}$ orbital occupation
(negative $P_{\rm e_g}$) in the electron-doped regime, compared to the
case of the tetragonally distorted cells.  Due to the released
symmetry constraint, the doped electrons can preferentially occupy the
Ni(1) sites and reduce the energy cost.  The smaller $d_{\rm
  Ni-O_{ap}} / d_{\rm Ni-O_{pl}}$ for Ni(1) (or increased $d_{\rm
  Ni-O_{pl}}$) indicates that additional electrons occupy the
$d_{x^2-y^2}$ orbital.  The Jahn-Teller type NiO$_6$ octahedral
distortion is observed in the undoped regime $U_{\rm eff}\leq$ 4
eV. This type of distortion has not been reported in previous
first-principles calculation studies in which the smaller value of
$U_{\rm eff}\approx$ 3 eV was used \cite{Chakhalian,Pentcheva}. Since
the Ni-$e_{\rm g}$ orbital is mainly occupied by one electron in the
undoped regime, the Jahn-Teller distortion can be a plausible way to
lower the electron correlation energy when $U_{\rm eff}$ is strong,
while it has not been reported in experimental studies\cite{Tung}. 
The degree of Jahn-Teller distortion at $U_{\rm eff}=$ 5 eV is shown
in Fig.~\ref{fig:rot_local}(d) as a function of doping.

\begin{figure}
  \centering
  \includegraphics[width=0.45\textwidth]{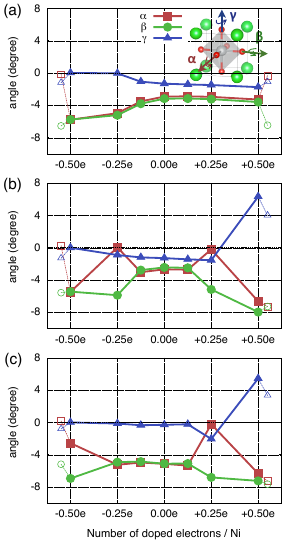}
  \caption{(Color online) The calculated rotation angles for (a)
    $U_{\rm eff}=$ 0, (b) 3, and (c) 5 eV as a function of doping.
    The angle $\alpha, \beta, \gamma$ are defined in the inset of (a).
    Following the Glazer notation, the positive and negative signs
    mean the ferro- and antiferro-distortive rotation, respectively.
  }
  \label{fig:rot_ang}
\end{figure}

The rotation of NiO$_6$ and TiO$_6$ octahedra also exhibits an
interesting pattern in response to the charge doping.
Fig.~\ref{fig:rot_ang} shows the calculated NiO$_6$ rotation angles as
a function of electron doping. At $U_{\rm eff}=$ 0 eV, the
antiferro-distortive angle $\gamma$ (see the inset of
Fig.~\ref{fig:rot_ang}(a)) is gradually increased from $0^{\circ}$ to
$-2^{\circ}$, while the angle $\alpha$ and $\beta$ are decreased from
$-6^{\circ}$ to $-3^{\circ}$.  Interestingly, this behavior is similar
to the structural change caused by biaxial compressive strain in
LaAlO$_3$, in which the lattice constant change is attributed solely
to the strain (without doping) while the size of the AlO$_6$ octahedra
remains basically unchanged\cite{Hatt}.  
In our system, the $d_{\rm Ni-O_{ap}} / d_{\rm Ni-O_{pl}}$ ratio 
increases as electrons are doped to the system, so the system should
feel an `effective' decrease of the c-axis constant, or equivalently,
the biaxial tensile strain if the c-axis constant had been fixed.
Instead, the c-axis lattice constant is increased under the electron
doping, enough to overcompensate the elongation of the NiO$_6$
octahedral. So the system feels an effective in-plane compressive strain
and the angle $\gamma$ increases slightly while $\alpha$ and $\beta$
reduce. 
Another point to mention is that the
rotation angle of the LaSr$_3$- and La$_3$Sr-cell show the same 
($a^-b^0c^0$) type pattern, which might be due to the arrangement 
of La$^{3+}$ and Sr$^{2+}$ cations and the resulting internal electric 
field along the c-direction.
Fig.~\ref{fig:rot_ang}(b) presents the
results for $U_{\rm eff}=$ 3 eV.  In the hole-doped and undoped regime
the results are similar with those at $U_{\rm eff}=$ 0 eV, except for
the case of $-$0.25$e$ doping where ($a^-b^0c^0$)-like rotation is
stabilized.  In the electron-doped regime, on the other hand, the
system changes from the ferrimagnetic to the half-metallic phase and
the rotation pattern also changes drastically to an ($a^-a^-c^+$)-like
phase as the electron doping increases to +0.50$e$/Ni. Such behavior
is also observed in the electron-doped regime at $U_{\rm eff}=$ 5 eV,
indicating that the large $d_{\rm Ni-O_{ap}} / d_{\rm Ni-O_{pl}}$
ratio as well as the presence of the breathing distortion might result
in the significant change of the rotation angle in these regimes. The
rotation pattern in the undoped and weakly hole-doped regime at
$U_{\rm eff}=$ 5 eV is similar to those at $U_{\rm eff}=$ 0 and 3 eV,
but with enhanced ($a^-a^-c^0$)-like character.  The La$_3$Sr-cell
calculation gives the qualitatively same rotation pattern with the
+0.50$e$-cells at $U_{\rm eff}=$ 3 and 5 eV, since the rotation angle
is large enough to overcome the effect of local ionic configurations
in this regime.

\subsection{Electronic structure, orbital polarization, and magnetism}

\begin{figure}
  \centering
  \includegraphics[width=0.45\textwidth]{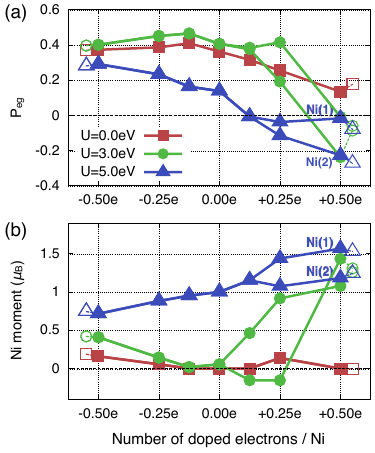}
  \caption{(Color online) (a) The calculated orbital polarization,
    $P_{e_{\rm g}}$, and (b) Ni-magnetic moment as a function of
    electron doping. Two different Ni sites, Ni(1) and Ni(2), at
    $U_{\rm eff}=$ 5 eV are marked in the plot (blue color).}
  \label{fig:rot_magnother}
\end{figure}

The calculated $P_{e_{\rm g}}$ as a function of doping in the fully
distorted calculations is summarized in
Fig.~\ref{fig:rot_magnother}(a), in which the decreasing tendency is
clearly observed. While this feature is similar to that in
Sec.~\ref{sec3}, the graph branches into two in the electron-doped
regime for $U_{\rm eff}=$ 3 and 5 eV.  In addition, $P_{e_{\rm g}}$
becomes smaller in the entire doping range because the released
structural constraint allows $d_{\rm Ni-O_{ap}}$ to be increased
by adopting octahedral rotation and tilting. The
$d_{3z^2-r^2}$ orbital occupation is favored accordingly. The
calculated PDOS provides additional information about the behavior
of the doped charge; as more electrons are introduced, an empty peak
of Ni(2) $d_{x^2-y^2}$ character emerges above the Fermi level while
the two $e_g$ orbitals of Ni(1) are equally occupied (see Appendix,
Fig.~\ref{figA:rot_pdos_eg}).  This feature can also be seen in
Fig.~\ref{fig:rot_local} where $d_{\rm Ni-O_{ap}} / d_{\rm Ni-O_{pl}}$
of Ni(1) reaches to unity in the electron-doped regime due to the
equal occupation of $d_{x^2-y^2}$ and $d_{3z^2-r^2}$ orbitals ({\it
  i.e.}  the vanishing $P_{e_{\rm g}}$ of Ni(1) in the electron-doped
regime at $U_{\rm eff}=$ 5 eV, see Fig.~\ref{fig:rot_magnother}.(a)).

The behavior of the Ni magnetic moment is also similar to that in the
tetragonally distorted cases (Fig.~\ref{fig:mag}(a)) for $U_{\rm
  eff}=$ 0 and 3 eV. However in this case, the half-metallic phase is
obtained in the calculation of +0.50$e$-cell with $U_{\rm eff}=$ 3 eV.
The antiferromagnetic phase disappears as shown in
Fig.~\ref{fig:PD}(b) due to the reduction of $P_{e_{\rm g}}$ in the
fully distorted cells.  The reduction of $P_{e_{\rm g}}$ also enlarges
the area of the half-metallic phase in $U_{\rm eff} \geq$ 4 eV. In the
electron-doped regime of the half-metallic phase, the magnetic moments
on the two inequivalent Ni atoms reflect the charge disproportionation
on the Ni sites.

\section{Discussion}
\label{sec5}


The doping simulated by the rigid band shift cannot be the same as what
happens in real experimental situations, such an oxygen vacancies. For
example, oxygen vacancies not only introduces effective electron
doping, but also distorts the local structure by disconnecting
the metal--oxygen--metal network.  However, we note that the two
different approaches, which incorporate the charge doping in our
calculations, produce consistent results regarding the change of the
electronic structure, orbital polarization and structural property.
Also, the orbital polarization is insensitive to some degrees of
structural difference.  These findings strongly suggest that the
overall conclusions presented in this study are quite relevant to
various doping situations in experiments, in spite of the limitation
of the simulation methods.

We emphasize that our results can provide useful information for
understanding texperiments. For example, the further distortion of
rotated oxygen octahedra caused by doping implies that additional oxygen
vacancies do not necessarily lead the system to be more metallic,
because further rotation can simultaneously make the system be less
metallic due to the enlarged effective $U/t$ parameter.  Also, our
prediction of polarization dependence as a function of electron doping
can be tested in experiments, for example by changing the oxygen
partial pressure in the pulsed laser deposition process.  Similar
doping-induced structural changes may happen in other oxide
superlattices with $d$-orbital degrees of freedom, such as the
LaTiO$_3$/LaAlO$_3$ system. One may also speculate the possibility of
rich phases from the LaNiO$_3$/LaTiO$_3$ superlattices, where the two
independent orbital degrees of freedom from the Ti-$t_{\rm 2g}$ and
Ni-$e_{\rm g}$ can interact through corner-sharing coupling of the
NiO$_6$ and TiO$_6$ octahedra.  Such materials as well as other
relevant systems may have substantial importance and deserve further
theoretical investigation.

\section{Summary}
\label{sec6}

The effect of charge doping on the electronic, orbital and structural
properties in LaNiO$_3$/SrTiO$_3$ has been investigated using
first-principles density functional theory calculation, in which doping
was simulated with two different methods, namely, rigid band shift and
the cation-substituted calculation.  The results clearly show the
systematic dependence of these physical properties on doping.  As more
electrons are introduced, the orbital polarization is reduced and the
Ni to apical oxygen distance significantly increases.  These features
are found in both tetragonally distorted and the fully distorted
results.  Remarkably, the rotation angles of the NiO$_6$/TiO$_6$
octahedra are also found to strongly depend on doping.
Our results suggest a possible way to control orbital and
structural property by means of charge doping, and provide useful
information to understand the experimental results under various doping
situations, such as oxygen vacancies.

\section{Acknowledgments}
M.J. Han thanks Michel van Veenendaal for fruitful discussion. This
work was supported by Basic Science Research Program through the
National Research Foundation of Korea(NRF) funded by the Ministry of
Education(2014R1A1A2057202).  H.-S. Kim was supported by Basic Science
Research Program through the National Research Foundation of Korea
(NRF) funded by the Ministry of Education (Grant
No. 2013R1A6A3A01064947). The computational resources were supported by
the National Institute of Supercomputing and Networking / Korea
Institute of Science and Technology Information including technical
support (KSC-2014-C3-050).


\newpage
\newpage

\appendix
\section{Projected density of states (PDOS)}

In this Appendix, we present the electronic structure change
as a function of doping, which provides further information
for understanding the doping dependence of our system, which
is closely related to the other physical quantities discussed above.

\begin{figure*}[h!]
   \centering
   \includegraphics[width=0.83\textwidth]{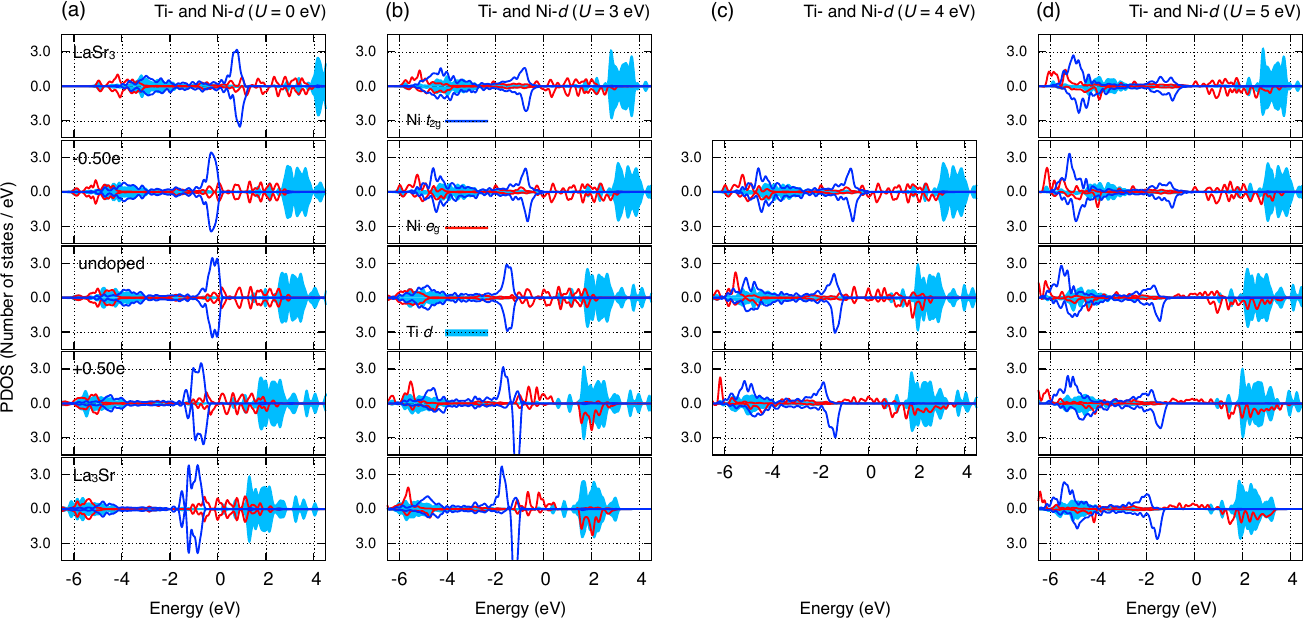}
   \caption{(Color online)
    The change of Ni-$d$ and Ti-$d$ PDOS as a function of doping at (a) $U_{\rm eff}=$ 0, 
    (b) 3, (c) 4, and (d) 5 eV in the tetragonally distorted calculations.
    Ni-$t_{\rm 2g}$, Ni-$e_{\rm g}$, and Ti-$d$-orbitals are represented by
    blue, red, and light blue lines, respectively.
    The first and last rows show the results from the LaSr$_3$- and La$_3$Sr-cell calculations,
    respectively.
    The second, third, and fourth rows correspond to the rigid band shift calculation 
    with $-$0.50$e$, 0.00$e$, and +0.50$e$ doping, respectively. 
   }
   \label{figA:pdos_d}
 \end{figure*}

\begin{figure*}[h!]
   \centering
   \includegraphics[width=0.83\textwidth]{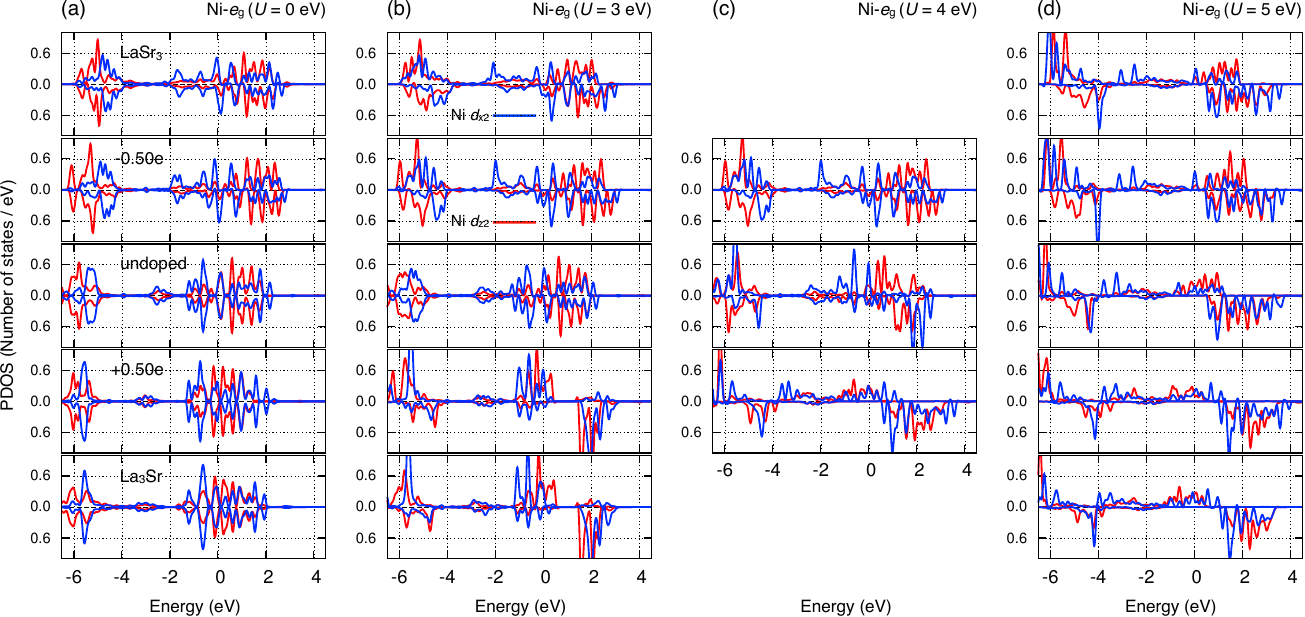}
   \caption{(Color online)
    The change of Ni-$e_{\rm g}$ PDOS as a function of doping at (a) $U_{\rm eff}=$ 0, 
    (b) 3, (c) 4, and (d) 5 eV in the tetragonally distorted calculations.
    Ni-$d_{x^2-y^2}$ and $d_{3z^2-r^2}$ states are represented by blue and red lines, respectively.
   }
   \label{figA:pdos_eg}
 \end{figure*}

\begin{figure*}[h!]
   \centering
   \includegraphics[width=1.0\textwidth]{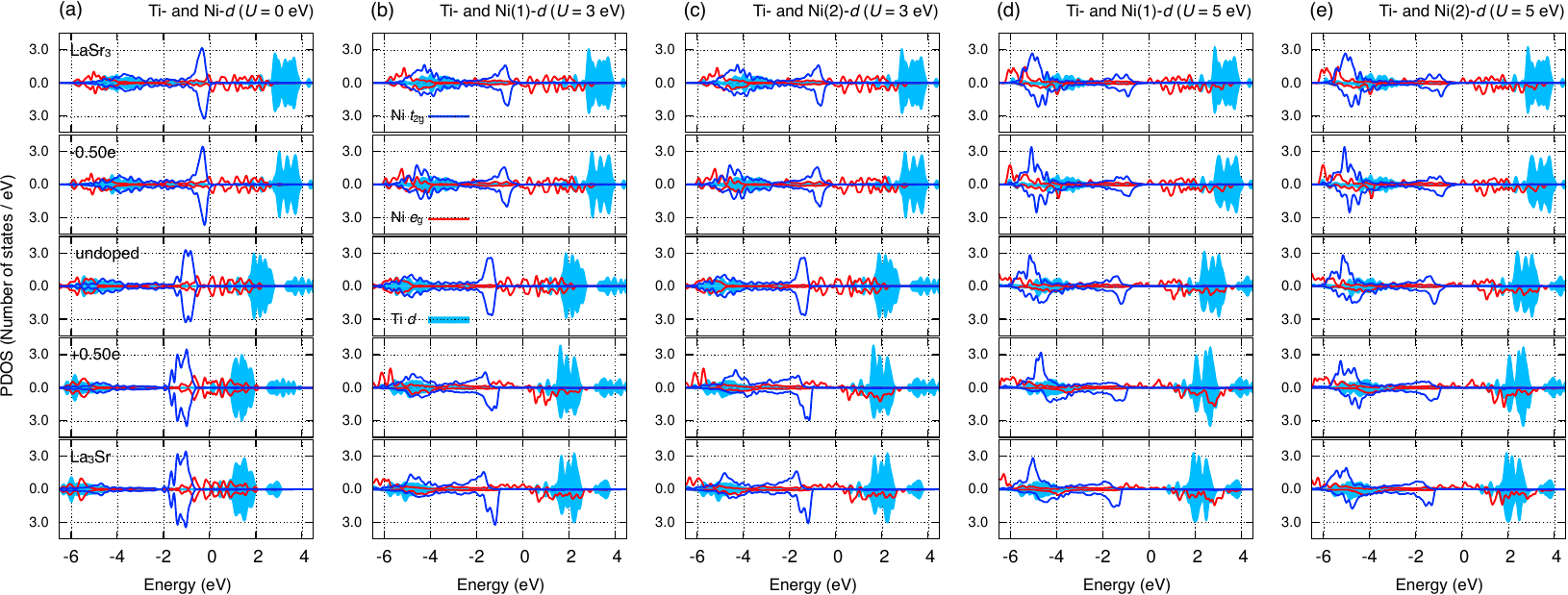}
   \caption{(Color online) 
    The change of Ni-$d$ and Ti-$d$ PDOS as a function of doping at (a) $U_{\rm eff}=$ 0, 
    (b) 3, (c) 4, and (d) 5 eV in the fully distorted calculations.
   }
   \label{figA:rot_pdos_d}
 \end{figure*}

\begin{figure*}[h!]
   \centering
   \includegraphics[width=1.0\textwidth]{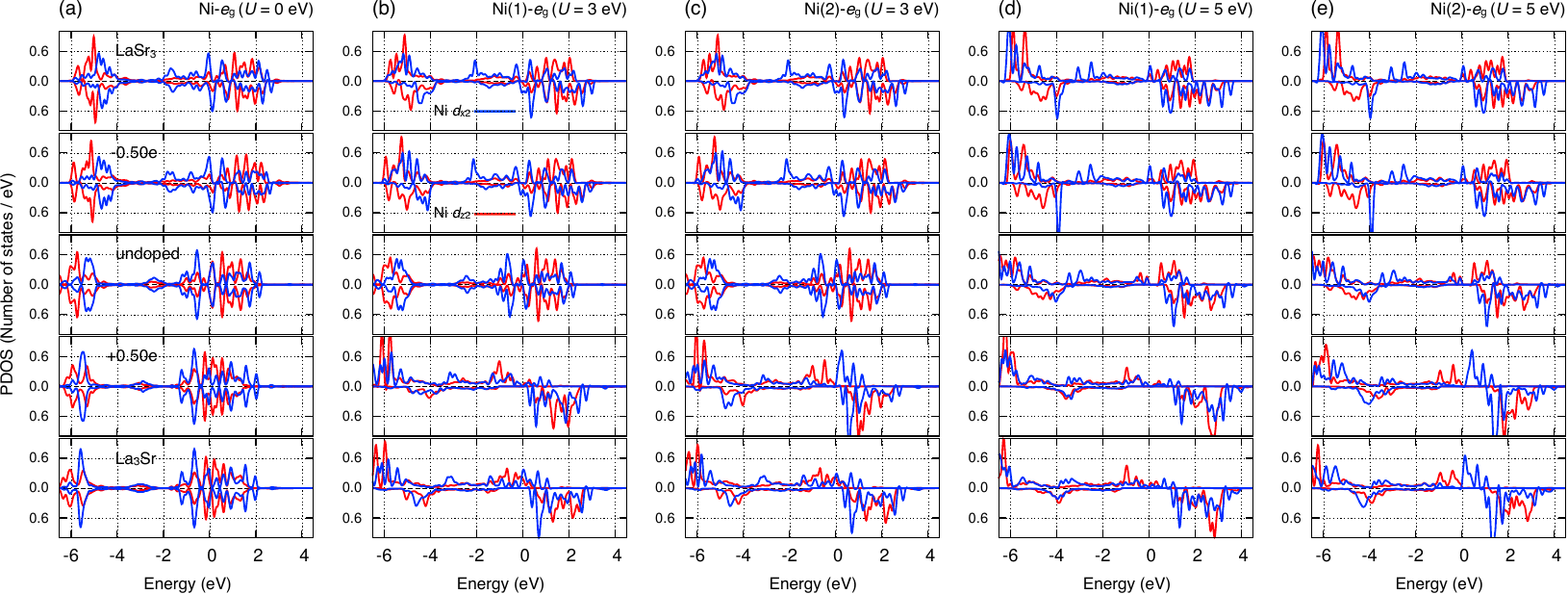}
   \caption{(Color online) 
    The change of Ni-$e_{\rm g}$ PDOS as a function of doping at (a) $U_{\rm eff}=$ 0, 
    (b) 3, (c) 4, and (d) 5 eV in the fully distorted calculations.
   }
   \label{figA:rot_pdos_eg}
 \end{figure*}

Figure \ref{figA:pdos_d} and \ref{figA:pdos_eg} shows the PDOS in the 
tetragonally distorted cells. First of all, we note that
the electronic structure difference between rigid band shift and
LaSr$_3$/La$_3$Sr-cell calculation is not significant.
This point also holds for the results of the fully distorted cells,
as shown in Fig.~\ref{figA:rot_pdos_d} and Fig.~\ref{figA:rot_pdos_eg}.
It is therefore consistent with our finding that two different approaches predict
the same features regarding the orbital occupation and structural properties as
discussed above. 
The overall shape of the Ni-$e_{\rm g}$ PDOS in the weakly correlated regime 
is actually consistent with the schematic picture in Fig.~\ref{fig:atomdisp}(a), 
vindicating our discussion in Sec.~\ref{sec3}.B based on this picture. 

Fig.~\ref{figA:rot_pdos_d} and \ref{figA:rot_pdos_eg} show the PDOS
from the structures with octahedral rotations in which the 
Jahn-Teller or breathing type distortion is incorporated (main results 
discussed in Sec. \ref{sec4}). 
Compared to the rotation-free results, due to the rotations, 
additional splittings are introduced in the Ni-$e_{\rm g}$ states
as clearly shown in Fig.~\ref{figA:rot_pdos_eg}. Also, in the presence of $U_{\rm eff}$,
electron doping induces charge disproportionation
(see the fourth and fifth rows of Fig. \ref{figA:rot_pdos_eg}(b)-(e)).
From the PDOS results and the data in Fig.~\ref{fig:rot_magnother}, 
one can consider that Ni(1) and Ni(2) (Fig.~\ref{figA:rot_pdos_eg}(b), (d) and (c), (e)) 
behave like as the $d^8$- and $d^7$-configuration in the electron-doped regime, respectively.

\newpage
\newpage

\newpage
\newpage

\end{document}